\documentstyle[pre,aps,epsfig,eqsecnum]{revtex}

\begin{document}
\draft

\title{Simulated emergence of cyclic sexual-asexual reproduction}
\author{J.S. S\'a Martins $^{\dag}$ and A. Racco $^{*}$}
\address{$\dag$ Colorado Center for Chaos and Complexity, CIRES, CB 216,  
University of Colorado, Boulder, Colorado 80309}
\address{$*$ Instituto de F\'{\i}sica, Universidade Federal 
Fluminense \\
Avenida Litor\^anea s/n, \,24210-340 , Niter\'oi, Rio de Janeiro, Brazil}

\maketitle

\begin{abstract}
Motivated by the cyclic pattern of reproductive regimes observed in some 
species of green flies (``{\it aphids}''), we simulate the evolution of
a population enduring harsh seasonal conditions for survival. The 
reproductive regime of each female is also seasonal in principle and 
genetically acquired, and can mutate for each newborn with some small
probability. The results show a sharp transition at a critical value
of the survival probability in the winter, between a reproductive
regime in the fall that is predominantly sexual, for low values of
this probability, or asexual, for high values.   
\end{abstract}

\section{Introduction}
The question of why sexual reproduction prevails in the immense majority 
of metazoan species still defies a complete and satisfactory answer -
for a recent review we direct the reader to Reference \cite{science}. 
A number of different theories have been put forth to try to explain this 
puzzle, but all have so far fallen short of becoming a paradigm in the field. 
One of the major difficulties one finds in this research is the lack of 
convincing observational data with which to compare one's theoretical effort. 
This is especially critical when addressing the onset of sexual reproduction 
itself, and in this case we can only speculate about this origin, but cannot 
test these speculations. A more promising line of research lies in analyzing 
the maintenance of sexual reproduction against the establishment of an 
apparently more efficient asexual variety. This efficiency stems from the 
absence of the necessity to generate males, which by itself alone should give 
the asexual variety the upper hand against a competing sexual alternative 
\cite{jms,gabriel}. 
Explaining the evolutionary choice favoring sexual reproduction in spite of 
its shortcomings has been the long-term goal for a number of researchers in 
this area.

It is not surprising that computer simulations have become a major tool in 
this field over the last few years. With the introduction of a number of 
simple simulational models, theoretical ideas began to be put to a ``live'' 
test on virtual populations. These models encompass dynamic rules that mimic 
the action of natural evolution in a variety of ways, and create a 
manageable representation of the conflict between selection and mutation, 
believed to be the driving evolutionary force. Among the models that have 
been used, we make special mention of the bit-string class, using 
Monte Carlo techniques \cite{tjpp,pekalski,erzan,dasgupta}. They seem 
particularly well suited to represent the microscopic dynamics of
genetic evolution and allow for a very efficient coding on personal 
workstations; recent and comprehensive reviews of a popular model of
this class, the Penna model, are available in the literature 
\cite{evolution,racco,school}.

Penna's model has in fact been recently used to address the foregoing problem. 
The greater genetic diversity generated by sexual reproduction and its 
consequences, such as survival after catastrophes \cite{why} and
winning when in competition against an asexual variety in the presence
of genetically coupled parasitism \cite{redq}, were already considered
in its context. A recent simulation including the action of
pleiotropic genes, i.e. those that have multiple effects, showed that
in a region of its parameter space sexual reproduction generates
larger equilibrium populations than its asexual counterparts when
evolving independently \cite{stauffer}. We report here on the use of a
modified Penna model in a different, though closely related, problem.

The problem in question here is that of the holocyclic behaviour of some 
species. By this is meant the capability shown by these species of 
seasonally alternating their reproductive regimes. Of particular interest 
in this class of species are the aphids, or green flies. 

Aphids are a group of about $4 \ 400$ species of small insects that feed on 
the phloem fluid of plants \cite{blackman,web}. Many species of aphids
attack important agricultural crops, and are therefore of major
economic importance, and display complex life cycles. They are one of
the few groups of animals that undergo cyclical parthenogenesis, that
is, the alternation of a varying number of asexual generations 
(parthenogenesis) with a single generation of sexual reproduction. All
asexual generations are entirely female. Species that produce both
sexual and asexual morphs are called holocyclic, as opposed to those 
anholocyclic ones that undergo obligate parthenogenesis. In a holocyclic 
life cycle both asexual and sexual morphs are produced at different times of 
the year. Males are present only in one part of the life cycle. They are 
usually produced only in the autumn by parthenogenetic
females. Females reproducing sexually produce just one egg, that in 
temperate regions overwinters and hatches only in the spring.

The purpose of the present work is to study the emergence of an holocyclic 
life cycle through a simple evolutionary model, with dynamics designed to 
mimic the main features of the relation between the aphids and their 
environment. To summarize, we show that this simple model has a first order 
holocyclic transition dependent on one parameter. On one side of the 
transition there is no alternation of reproductive regime and the species is 
asexual; on the other, the holocyclic order is chosen by evolution as a means 
of insuring the survival of the species.

This paper is organized as follows: in the next section we present the model 
that was used to simulate an aphid-like species; our results are discussed 
in Section III, and we conclude in Section IV.

\section{The Model}
One of the main features of the environmental impact on the aphids is their 
great susceptibility to winter conditions. The majority of the population 
cannot survive in very cold conditions. On the other hand, aphid eggs 
generated by sexual reproduction resist harsh weather conditions. They 
overwinter and hatch only when spring comes. These are the elements to be 
introduced into a model that intends to test the hypothesis that the onset of 
holocyclic order is a direct consequence of evolution dynamics acting on 
cold-susceptible species.

The model we worked with is an adaptation of the Penna model for 
age-structured populations, to which we added the features described above. 
Since the phenomenon in question does not depend on the age structure, this 
model is not the only possible choice. Nevertheless, it has a very simple 
representation of the conflict between mutations and selection, which is 
central to any evolutionary problem, together with a built-in mechanism 
that decreases the survival probability with age, that other models have to 
introduce in an {\it ad hoc} way. We will briefly describe the Penna model 
in its mixed sexual-meiotic parthenogenesis version. A similar version of 
this model - but without the specific aphid-type dynamics - was used in 
Ref. \cite{redq}.

The genome of each (diploid) organism is represented by two computer
words. In each word, a bit set to one at a position (``{\it locus}'')
corresponds to a deleterious mutation - a ``perfect'' strand would be
composed solely of zeros. The reader should note that we don't allow
in this context for a beneficial mutation in the wildtype genome, that
would eventually lead us to the problem of speciation. The effect of a
harmful mutation may be felt by the individual at all ages equal to or
above the numerical order of that {\it locus} in the word. As an
example, a bit set to one at the second position of one of the
bit-strings means that a harmful effect may become present in the life
history of the organism to which it corresponds after it has lived for
two time periods. The diploid character of the genome is related to
the effectiveness of the mutations. A mutation in a position of one of
the strands is felt as harmful either because of homozygose or because
of dominance. For the former, a mutation must be present in both
strings at the same position to be effective. The concept of dominance on
the other hand relates to {\it loci} in the genome in which a mutation in
just one strand is enough to make it affect the life of the organism. The life
span of an individual is controlled by the amount of effective mutations
active at any instant in time. This number must be smaller than a
specified threshold to keep the individual alive; it dies as soon as this
limit is reached. 

Reproduction is modeled by the introduction of new genomes in the
population. Each female becomes reproductive after having reached a
minimum age, after which she generates a fixed number of offspring at the
completion of each period of life. The meiotic cycle is represented by
the generation of a single-stranded cell out of the diploid genome. To do
so, each string of the parent genome is cut at a randomly selected
position, the same for both strings, and the left part of one is combined
with the right part of the other, thus generating two new combinations of
the original genes. The selection of one of these completes the formation
of the haploid gamete coming from the mother. 

The difference between sexual and parthenogenetic reproduction appears at
this stage. For the first, a male is selected in the population and
undergoes the same meiotic cycle, generating a second haploid gamete out
of his genome. The two gametes, one from each parent, are now combined to
form the genome of the offspring. Each of its strands was formed out of a
different set of genes. 

For meiotic parthenogenesis, all genetic information of the offspring comes 
from a single parent. Its gamete is generated through the meiotic cycle 
described above and then cloned to compose an homozygous genome for
the offspring. For both regimes, the next stage of the reproduction
process is the introduction of $M$ independent mutations in the newly
generated genetic strands. In this kind of model it is normal to consider
only the possibility of harmful mutations, because of their overwhelming
majority in nature. If male offspring can be born, the gender of the newborn 
is then randomly selected, with equal probability for each sex. 

A last ingredient of the model is a logistic factor, called the Verhulst
factor, which accounts for the finite carrying capacity of the environment 
for this particular species. It introduces a mean-field probability of death 
for an individual, coming from causes with dynamics not included in the 
model, and for computer simulations has the benefit of limiting the size of 
populations to be dealt with. This factor is essentially the ratio between 
the actual population in any time step and a parameter of the model, 
traditionally and perhaps improperly called the carrying capacity of the 
environment. Because it is not connected to the quality of the individual's 
genome, the usage of this factor in a model of evolutionary dynamics has its 
shortcomings \cite{cebrat}, and one has to make sure that the outcome of the 
simulations is not biased by the particular choice of strategy for its 
implementation.

The passage of time is represented by the reading of a new {\it locus} in
the genome of each individual in the population, and the increase of
its age by one. After having accounted for the selection pressure of a
limiting number of effective (harmful) mutations and the random action of 
the Verhulst dagger, females that have reached the minimum age for 
reproduction generate a number of offspring.
The simulation runs for a pre-specified number of time steps, at the end
of which averages are taken over the population(s).

In the mixed sexual-parthenogenesis version, each female in the population 
has a season-dependent reproductive mode, which is also genetically 
inherited. Four bits in the genome represent the reproductive mode for each 
season. A bit set to $0$ ($1$) in a position indicates that at the season 
associated to that particular position this female will generate offspring 
through sex (meiotic parthenogenesis). This reproductive pattern is passed 
on to the female offspring. Mutations can also occur on this pattern, with 
some probability. One out of the four seasons is randomly selected and the 
reproduction mode for that season is switched.

The duration of the seasons is set to some number of time periods. At each 
time step, the season to which it corresponds is calculated. In the fall, 
females that reproduce by parthenogenesis can generate both male and female 
offspring. In the winter, the carrying capacity of the environment is reduced 
by some factor, called here the compression, or burdening, factor. 
Individuals with age zero that were generated by sexual reproduction - and are 
protected by the eggshell - suffer only from the normal Verhulst factor, 
representing the action of predators, and do not age; they only hatch in the 
spring. All the others, irrespective of gender, age or reproductive pattern, 
suffer the action of the enhanced Verhulst factor. The fertility of females 
also depends on the reproductive regime, and we fixed it to $1$ offspring 
per time period for sexual reproduction.

Averages are taken of the total population, number of females with each 
reproductive pattern - both total and only the ones mature for reproduction - 
for each season. These numbers are normalized to the normal carrying capacity. 

The runs from which our data was collected had as common parameters $12$ 
aphid time periods for season duration, thus fixing our time scale as 
roughly a week per time step, a probability for reproductive pattern 
switching of $0.01$, and an equal selection of each gender for 
asexually-generated offspring in the fall. In all cases, the initial 
population had a season-independent sexual reproductive pattern, and was 
composed of the same number of individuals in each gender.

For the values of the standard parameters of the Penna model we used:

\noindent 1) Carrying capacity: $C = 800 \ 000$;

\noindent 2) Threshold of harmful mutations: $T = 3$;

\noindent 3) Minimum reproduction age: $R = 10$;

\noindent 4) Mutation rate: $M = 2$;
 
\noindent 5) Birth rate: $b = 10$ per female per time step for asexual 
reproduction;

\noindent 6) Initial population: $10 \ 000$ males and $10 \ 000$ females.

\section{Simulation results}
\begin{figure}[htb]
\centerline{\psfig{file=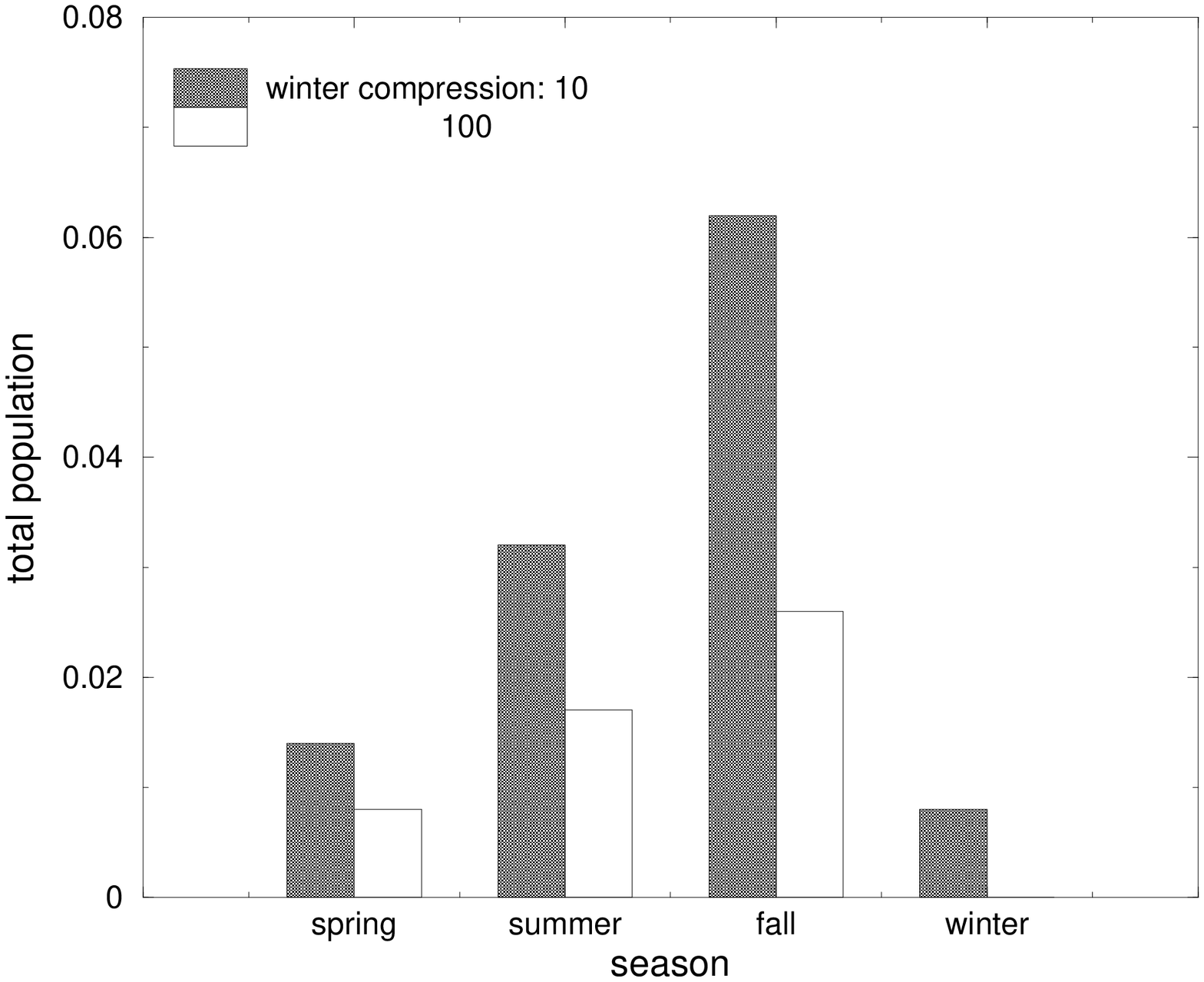, width=14cm, angle=270}}
\caption{The histogram shows the surviving population with age $>0$ in each 
season, as a dimensionless fraction of the carrying capacity parameter
$C$. The shaded bars represent data for a small winter compression 
factor of $10$ and the light bars data for a large factor of $100$. In this 
last case the population completely vanishes in winter time, except for the 
offspring of females that reproduced sexually in the fall - not represented 
in this graph.}
\label{popsea}
\end{figure}

We begin by showing the total population for each season in Figure 
\ref{popsea}. The figure for the population is normalized to the normal 
carrying capacity of the environment, one of the parameters of the model. 
Data is collected as an average over the last $960$ steps of a run of 
$50 \ 000$ Monte Carlo steps, for winter compression factors of $10$
and $100$. 
In both cases, the population increases from spring to fall, and decreases, as 
expected, in the winter season. For a compression factor large enough, the 
harsh conditions in this last season cause the population to vanish almost 
completely. The species does not disappears thanks to the remaining offspring 
of females that reproduced sexually during the last time step in the fall. 

\begin{figure}[htb]
\centerline{\psfig{figure=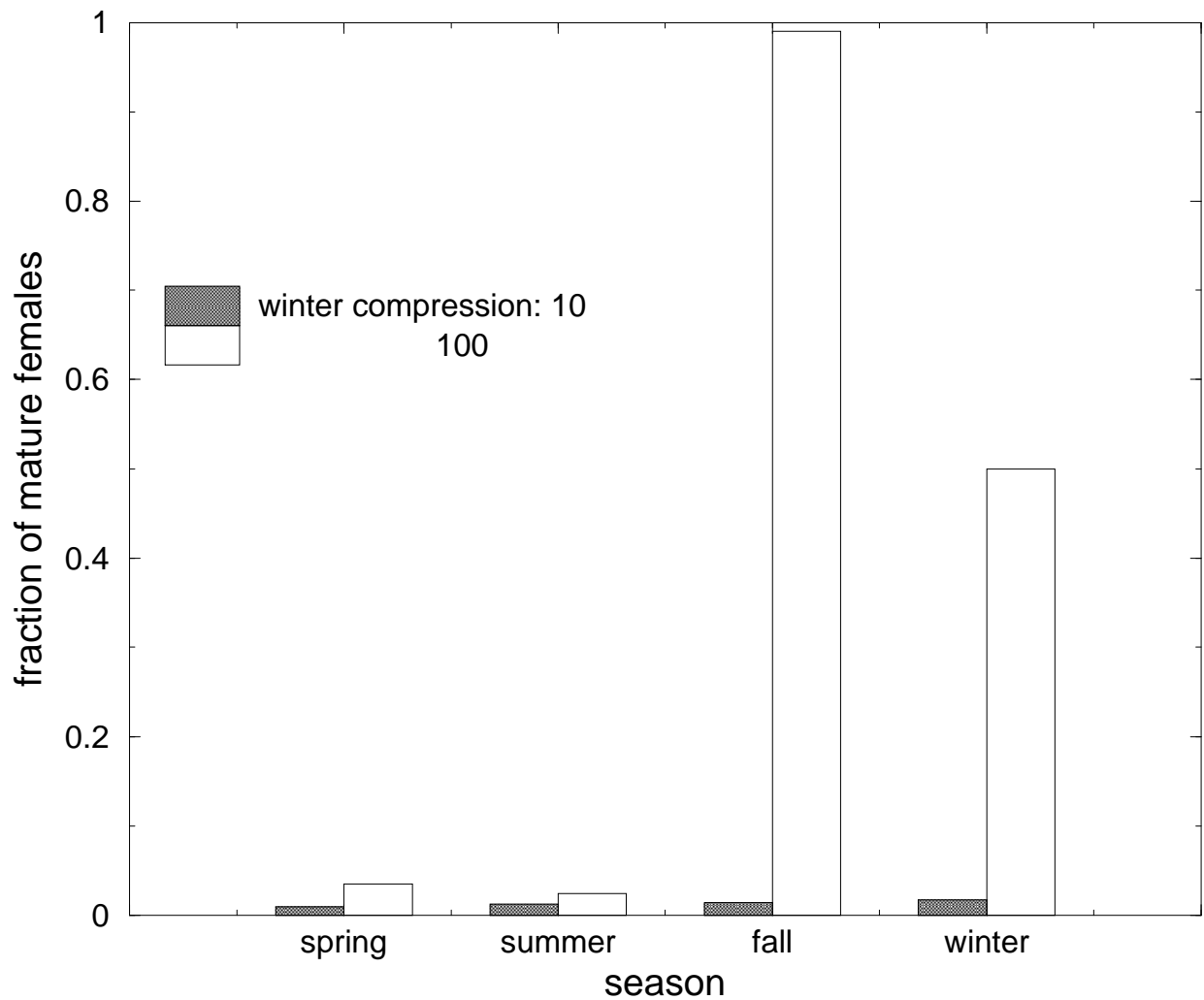, width=14cm, angle=270}}
\caption{The dimensionless fraction of mature females - those that have 
already reached the minimum age for reproduction - that reproduce sexually in 
each season is shown in this histogram. Again, shaded bars correspond to a 
compression factor of $10$, whereas light bars represent data for a factor of 
$100$. For the former case, females that reproduce sexually are present very 
marginally in the population. As for the latter, almost all females reproduce 
sexually in the fall. For the winter, the value of 0.5 simply reflects a 
non-biased random choice, since there are no females alive to reproduce.}
\label{repsex}
\end{figure}

Figure \ref{repsex} shows the fraction of mature females - those that have an 
age greater that the minimum reproduction age - that reproduce sexually in 
each season. If the winter has a mild effect on the effective carrying 
capacity, the evolution pressure favors reproduction through asexual 
meiotic parthenogenesis. Sexually-reproducing females are still present in the 
population, as a result of a small probability of back mutations from the 
prevailing asexual mode. The situation changes dramatically when the effects 
of the winter get stronger. If the corresponding compression factor is capable 
of bringing down the population in winter time to a small enough fraction of 
the normal carrying capacity, the reproductive regime in the fall switches to 
a largely predominant sexual variety. The data shown in this histogram 
correspond to situations that are way out of the transition region, and must 
be compared with Figure \ref{popsea}. In particular, the data shown there for 
the winter season determine what we see in the present histogram for the fall 
season.

The data for the winter season in Figure \ref{repsex} deserve a short comment. 
Since the mature female population in winter time has all but vanished for 
a compression factor of $100$, evolution exerts no pressure whatsoever
over the reproductive regime in this season. As a consequence, the 
distribution of females among the two regimes is random, and the
resulting fraction of sexually reproducing ones fluctuates around $0.5$.

\begin{figure}[htb]
\centerline{\psfig{file=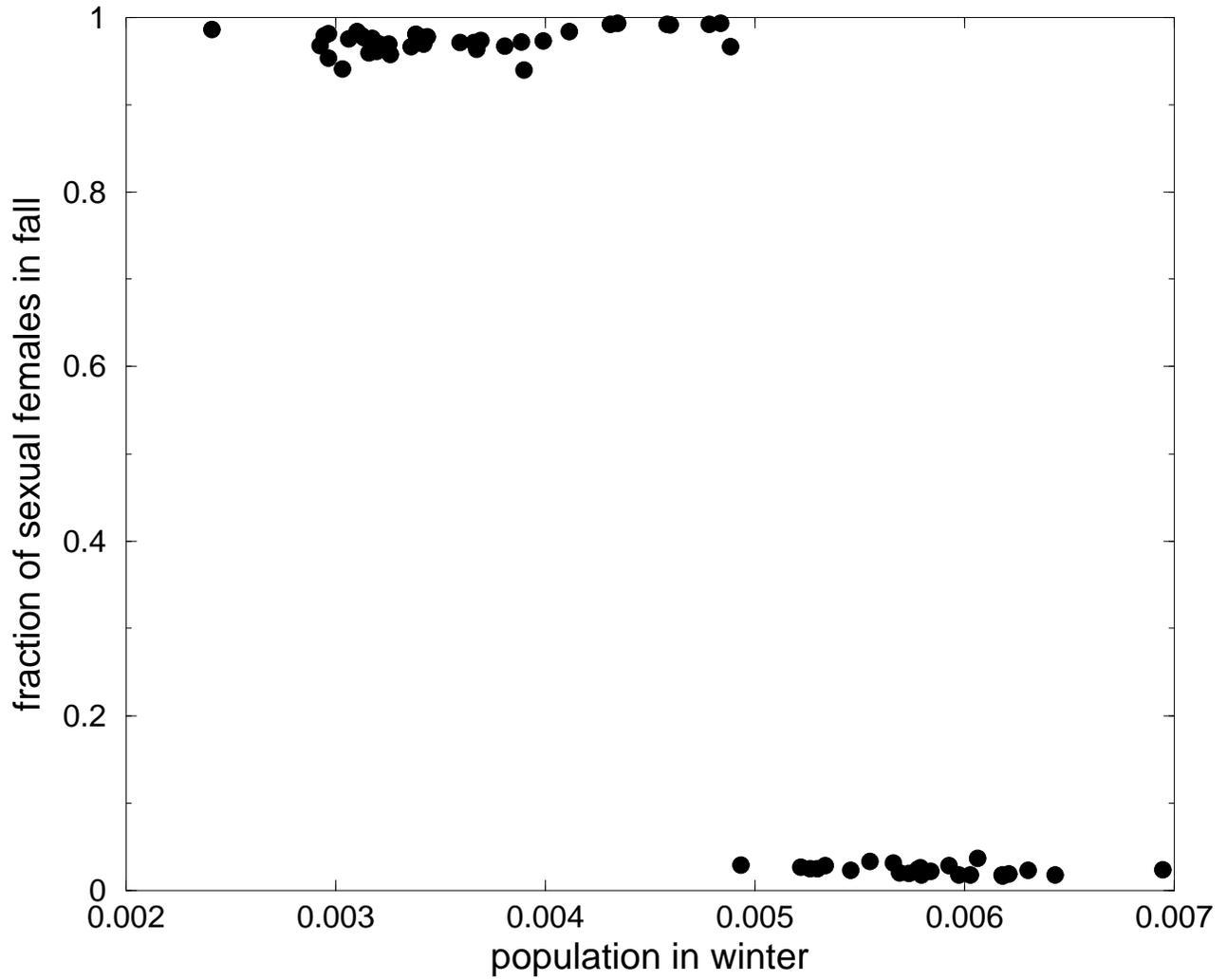, width=14cm, angle=270}}
\caption{The dimensionless fraction of sexually-reproducing females in the 
fall as a function of the surviving population in winter, expressed as a 
dimensionless fraction of the carrying capacity parameter $C$. It is 
clear that this surviving population is in fact the parameter that
controls the onset of the transition from a predominantly
parthenogenetic regime to a sexual one, as its value decreases.}
\label{ordpar}
\end{figure}

We proceed to characterize the transition between the anholocyclic and 
holocyclic regimes. The order parameter of this transition is clearly the 
fraction of sexually reproducing females in the fall season, which is $0$ in 
one regime and $1$ in the other. Figure \ref{ordpar} shows the behaviour of 
this parameter as a function of the surviving population in winter. A very 
sharp switch in regimes can be seen when this parameter is $0.0049$. Some of 
the runs near the transition region also had an unusually long relaxation 
time. These two characteristics point to the identification of a first order 
transition, and the long relaxation times can be associated to long-lived 
meta-stable states equivalent to super-cooled states in fluid transitions. 
A sharp transition of this kind has also been found in the study of the 
evolution of reproductive regimes in rapidly mutating ecologies \cite{redq}, 
and may well be an universal feature of living systems.
 
\begin{figure}[htb]
\centerline{\psfig{file=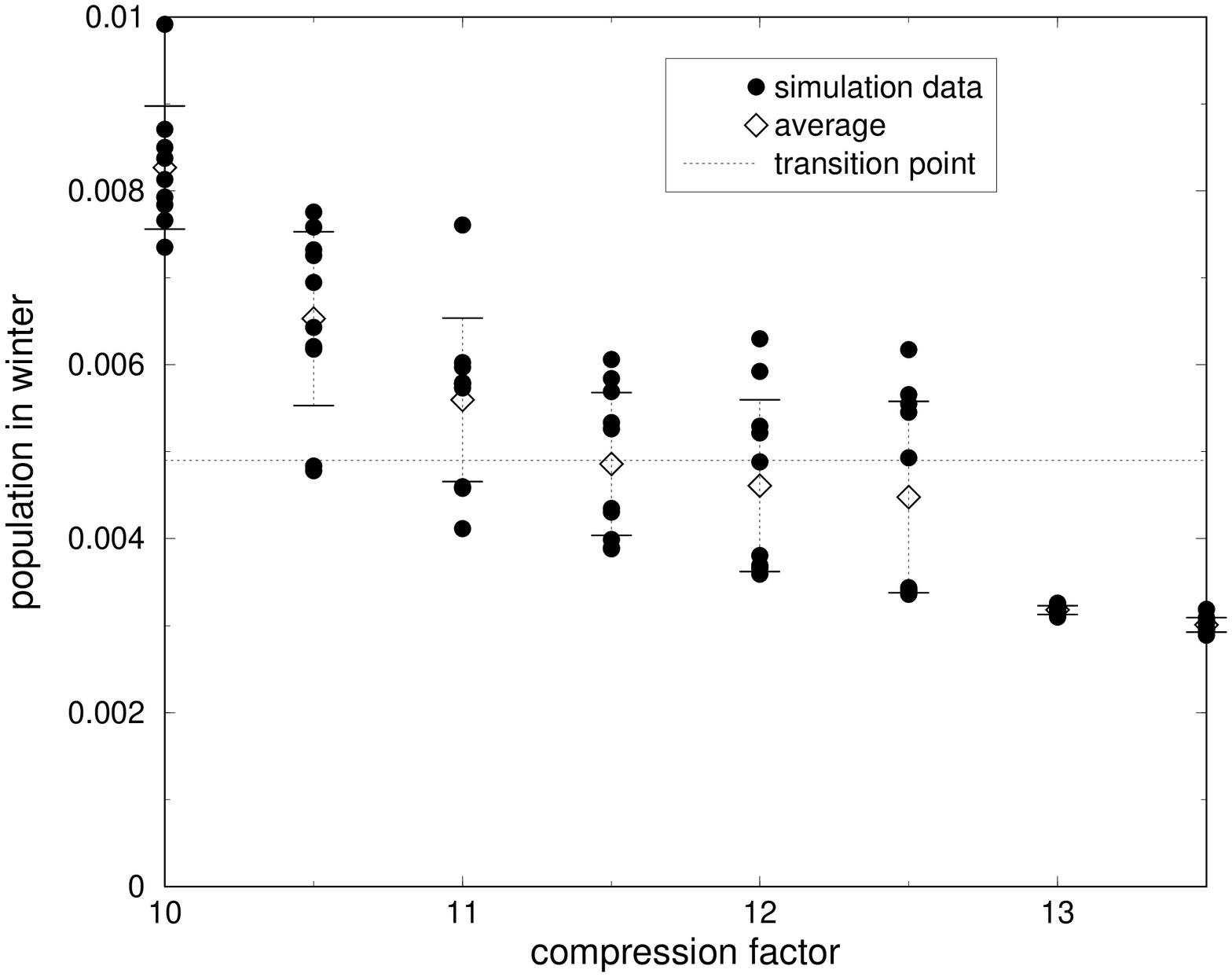, width=14cm, angle=270}}
\caption{The surviving population in winter time - expressed as a 
dimensionless fraction of the carrying capacity $C$ - as a function of 
the dimensionless compression factor. The transition region, standing for 
values of this last factor between $10.5$ and $12.5$, show a large dispersion 
for the final population. The dotted horizontal line represents the value of 
this population at the transition, $0.0049$. Also shown - as diamonds in the 
plot - are the averages for each compression factor, together with the 
standard deviations.}
\label{burden}
\end{figure}

The natural candidate in the model to act as a measure of the distance to the 
transition point appears at first to be the compression factor. But the 
simulation results show very strong fluctuations in the order parameter when 
this compression parameter is in the range $10.5$ - $12.5$. A closer 
examination of the outcome of these simulations clarifies this situation. In 
Figure \ref{burden} the final value of the population is shown as a function 
of this compression factor. Results are shown for $10$ different runs for each 
value of this parameter, lasting $200 \ 000$ Monte Carlo steps each. The 
dispersion of the final population for a compression factor in the above range 
causes an end result that can be in either side of the transition. Although 
not directly accessible as an input parameter to the simulations, the real 
control variable in this case is the surviving population in the winter, and 
Figure \ref{ordpar} is an eloquent support for this claim. Also shown in 
Figure \ref{burden} are the average values for the population in winter as a 
function of the compression factor, with error bars representing the standard 
deviation. The horizontal dotted line stands for the value of the population 
at the transition point, $0.0049$.

\section{Conclusions}
As was already shown in Ref. \cite{redq}, in a stable environment and
when no pleiotropic genes are present, the advantage of not having to
carry males gives asexual reproduction the upper hand in a direct
competition against sex. This is seen by the small fraction of females
that carry a sexual reproductive pattern in spring and summer shown in
Figure \ref{popsea}. And even when their offspring are given the
advantage of the eggshell protection in winter time, the switching of the 
dominant mode of reproduction in the fall only occurs for a very small 
surviving population in the winter. For large values of this surviving
population, the pattern of reproduction is asexual and 
season-independent, switching to a cyclic asexual-sexual
season-dependent one when the surviving population is smaller than some
threshold. The establishment of an holocyclic order in the population
derives thus from the advantages of overwintering under eggshell
protection, and has no direct relation to the genetic diversity
generated by sex. We could identify it as a first-order process driven
by the surviving population; this claim is supported by the sharp
transition observed when the fraction of sexually reproducing females
in the fall is considered as an order parameter for this transition.

\section*{Acknowledgments}
We are greatly indebted to F. Delmotte, who introduced us to specialized 
literature on the biology of aphids, and to our referee for a number
of suggestions that improved the clarity of our presentation and
updated our references. 
J.S.S.M.'s work is supported as a Visiting Fellow by CIRES, University 
of Colorado at Boulder, and A.R. acknowledges financial support by the 
Brazilian agency CNPq.

\newpage

\begin{thebibliography}{99} 

\bibitem{science}Science {\bf 281}, 1979-2008 (1998).

\bibitem{jms}J. Maynard Smith, {\it The Evolution of Sex}, 
(Cambridge University Press, Cambridge, England, 1978).

\bibitem{gabriel}E. Baake and W. Gabriel, in {\it Annual Reviews of 
Computational Physics}, vol. VII, pg. 203, Ed. D. Stauffer (World
Scientific, Singapore, 2000).

\bibitem{tjpp}T.J.P. Penna, J. Stat. Phys. {\bf 78}, 1629 (1995).

\bibitem{pekalski}A. Pekalski, Physica A {\bf 265}, 255 (1999).

\bibitem{erzan}B. \"Or\c cal, E. T\"uzel, V. Sevim, N. Jan, and A. Erzan,
  Int. J. Mod. Phys. C {\bf 11}, 973 (2000); E. T\"uzel, V. Sevim, and
  A. Erzan, e-prints cond-mat/0103010 and cond-mat/0101426 (2001).

\bibitem{dasgupta}S. Dasgupta, preprint (2001).

\bibitem{evolution}S. Moss de Oliveira, P.M.C. de Oliveira, and D. Stauffer, 
{\it Evolution, Money, War and Computers} (Teubner, Leipzig, Germany, 1999).

\bibitem{racco}T.J.P. Penna, A.Racco, and M.A. de Menezes, 
Comput. Phys.Commun. {\bf 121-122}, 108 (1999).

\bibitem{school}S. Moss de Oliveira, D. Alves, and J.S. S\'a Martins, 
Physica A {\bf 285}, 77 (2000).

\bibitem{why}J.S. S\'a Martins and S. Moss de Oliveira, 
Int. J. Mod. Phys. C {\bf 9}, 421 (1998).

\bibitem{redq}J.S. S\'a Martins, Phys.Rev. E {\bf 61}, R2212 (2000).

\bibitem{stauffer}J.S. S\'a Martins and D. Stauffer, Physica A {\bf
    294}, 191 (2001).

\bibitem{blackman}R.L. Blackman, Reproduction, cytogenetics and development; 
in {\it Aphids: Their biology, natural enemies, and control}, Vol 2A, 
Eds. A. K. Minks and P. Harrewijn, (Elsevier, Amsterdam, 1987).

\bibitem{web}A short introduction to the aphids for the laymen can be found 
at www.chu.cam.ac.uk/aphids/aphidomorpha.html.

\bibitem{cebrat}J.S. S\'a Martins and S. Cebrat, Theory Biosci. {\bf 119}, 
156 (2000).

\end{thebibliography}
\end{document}